\documentclass[conference]{IEEEtran}
\IEEEoverridecommandlockouts

\usepackage{tabularx,booktabs}
\usepackage{dingbat}
\usepackage{diagbox}
\newcolumntype{C}{>{\centering\arraybackslash}X}
\usepackage{cite}
\usepackage{amsmath,amssymb,amsfonts}
\usepackage{algorithmic}
\usepackage{makecell, longtable}

\usepackage{stfloats}

\usepackage{amsmath}
\usepackage[skip=1ex]{caption}
\usepackage{graphicx}
\usepackage{textcomp}
\usepackage{xcolor}
\def\BibTeX{{\rm B\kern-.05em{\sc i\kern-.025em b}\kern-.08em
    T\kern-.1667em\lower.7ex\hbox{E}\kern-.125emX}}
\begin{document}
\title{\LARGE \bf
Energy-Efficient Techniques for UAVs in Communication-based Applications
}
\author{\IEEEauthorblockN{1\textsuperscript{st} Anas Osman}
\IEEEauthorblockA{\textit{Department of Industrial Engineering} \\
\textit{University of Trento}\\
Trento, Italy \\
anas.osman@studenti.unitn.it}
\and
\IEEEauthorblockN{2\textsuperscript{nd} Morteza Alijani}
\IEEEauthorblockA{\textit{Department of Industrial Engineering} \\
\textit{University of Trento}\\
Trento, Italy \\
morteza.alijani@studenti.unitn.it}
}

\maketitle
\begin{abstract}
Unmanned Aerial Vehicles (UAVs), which are at the forefront of cutting-edge technology, have unmatched potential for pioneering applications in a wide range of disciplines. In particular, in the field of cognitive radio (CR), which is a key aspect in the implementation of the new 5G telecommunication technology. The integration between the drone and CR consolidates the drone's capabilities at the heart of the remarkably promising Internet-of-Things (IoT) technology supported by CR. The highly dynamic network topologies, weakly networked communication links, reliable line-of-sight (LOS) communication links, and orbital or flight paths are characteristic features of UAV communication compared to terrestrial wireless networks. Nevertheless, the implementation of this system is constrained by several severe challenges, such as energy efficiency, battery power limitation, spectrum handover, propagation channel modeling, routing protocols, security policy, and delay setbacks. In this paper, we consider the impact of energy scarcity faced by the UAV in various CR applications. We also analyze the impact of energy scarcity on communication-based applications and present the general problem of battery limitation. Finally, we give an overview and comparison between recent solutions proposed by researchers both in the field of communication and based on batteries, and consider possible future directions according to the state of the art, such as novel Graph Signal Processing (GSP) and machine learning (ML).
\end{abstract}
\section{Introduction}
Nowadays, the growing number of devices connected to the Internet has fueled the popularity of the ever-expanding Internet of Things (IoT) applications. On the other hand, the rapid development of the technology has exploited the vulnerabilities of the algorithms and techniques currently used, especially in the field of communication. Namely, the 5G service as one of the most anticipated mobile communication networks that has invigorated the industry and changed the network infrastructure used. The 5G service is expected to support a relatively high number of devices, which increases the speculation about the scarcity of spectrum, including higher (mmWave) and lower frequency spectrum, which is essentially one of the major challenges for applications related to cognitive radio [1].\\
In the last decade, Cognitive Radio has been widely regarded as the key to solving the increasingly scarce spectrum. Despite the high research effort, this technology still suffers from limitations such as multiple fading and shadowing [2]. Moreover, the current network infrastructure is not comprehensive enough in remote areas where building a terrestrial network is considered economically impossible [3]. In contrast, Unmanned Aerial Vehicles (UAVs), which are at the forefront of aerial technologies, is a promising solution to current network limitations. Figure 1, shows how UAVs are being integrated as part of the 5G network and next generation wireless networks as they can be automatically deployed in many wireless communication applications and leverage the capabilities of UAVs to overcome the current limitations. It is known that UAVs operate autonomously as they can be controlled using a remote control or from a control center [4]. They have proven to be effective in many applications such as traffic monitoring [5], surveillance [6], search and rescue [7], package delivery [8], disaster control [9], spectrum acquisition and modulation [10], fire control, etc. [11]. On the other hand, the integration of UAVs with cognitive radio as an intelligent system to detect actual occupancy in specific frequency bands may be beneficial for future wireless networks, by identifying the primary users (PUs) and dynamically shifting the parameter to allow access to unlicensed users while the spectrum is free. Furthermore, the combination of CR with UAVs has been shown to overcome existing problems such as dynamic spectrum access (as CR reduces interference with other signals and improves overall performance), excessive power consumption and delays [12].
For example, with respect to multiple fading and shadowing, which are the main problems associated with CR applications, UAVs have the ability to obtain a higher strength signal in the fading and shadowing scenario [2]. The authors in [2] proposed a UAV-based CR to improve spectrum detection performance and to access the unoccupied spectrum. They divided the process into two stages, including the formulation of spectrum sensing efficiency as an optimization problem and virtual cooperative spectrum detection. The simulation results showed the better UAV transmission and consequently higher performance in terms of spectrum sensing in deep fading channel [2]. Although integration between the UAV and CR has been introduced as an outstanding solution for dealing with the limitations of spectrum scarcity, this approach still suffers from many challenges and there is room for improvement. Figure 2, illustrates the important constraints associated with using the UAVs as part of wireless communications. The primary aspiration of this study is to limit energy consumption, which is considered as a fixed problem since typical UAVs are limited by battery size to fulfill the aerodynamic requirements, which leads to a definite energy shortage problem. In addition, the integration of a CR receiver on-board leads to a dramatic increase in energy consumption, resulting in a relatively high downtime for recharging and a noticeably shorter flight time. Furthermore, the energy consumed by the communication platforms and other processing functions has a direct impact on the lifetime of the battery-powered flight units in this structure [13]. The increase in energy consumption is significantly related to the spectrum hand-off, of which CR is an integral part, as it constantly checks PU for intervention to interrupt the transmission by SU. To counteract the battery limitations in UAV-based CR systems, the authors have proposed many techniques that have advantages and disadvantages. The main contribution of this paper is to analyze the current solutions and to provide comparative views of these approaches.\\
The paper is organised as follows: Section II is dedicated to the integration of UAVs with CR applications, related topics and the detailed overview table that classifies the proposed solutions according to the state of the art. Battery-based energy management, including challenges and potential techniques for energy efficiency management (EE), is described in detail in section III. Section IV presents the future roadmap, including potential directions such as novel graph signal processing (GSP) and machine learning (ML). Finally, Section V contains our conclusion.

\section{Integration of UAVs in Communication-based Applications, Issues, and Potential Solutions}
UAV networks have unique properties and characteristics ranging from slow-dynamic to dynamic, making them remarkably effective for communication-based applications [13,14]. Recent work in the areas of Mobile AD HOC Networks (MANETs) and Vehicular Ad HOC Networks (VANETs) takes advantage of the capabilities of UAV networks [12]. Nevertheless, the researchers consider this area as a major challenge due to the outstanding problems related to the implementation of UAV networks, as shown in Figure 2. As already mentioned, in this study we specifically examine the issue of energy conservation on the UAV-based communication networks. Recently, many researchers have proposed energy optimization techniques to improve flight duration and increase flight range. Most of the proposed methods are implemented by manipulating control algorithms and managing the resources allocated to avionics. This is because a physical change in battery size would require a special UAV design that differs from the conveniently available commercial drones. Most of the battery consumption is driven by the multiple motors that keep the drone hovering, so that the energy consumption of the network is relatively negligible [15-17]. In order to optimize energy consumption, the influence of both hardware (battery) and wireless communication on energy consumption must be considered [15,16]. Therefore, in this study, energy efficient techniques (EE) based on communication operations are discussed in section II and battery-based energy management (BBEM) approaches are analyzed in section III.
Furthermore, in this paper we classified the communication-based solutions into two parts according to the main issues of wireless communication structures, as the first category is dedicated to the optimization of measurement time and transmission power. In [17], for example, the authors examined the two main challenges that limit the integration of UAVs in a CR system and discussed in particular the energy optimization of a MUAV (Micro Unmanned Aerial Vehicle). Furthermore, they opted for a joint optimization approach to predict an optimized 3D location of the drone while maintaining the transmission power control at an optimal level, based on the Weber formulation in the sense that the MUAV is capable of detecting PU. The authors stated that the results obtained were similar to those obtained with the conventional Particle Swarm Optimization Algorithm (PSO). Similarly, the system tries to overcome the extensive energy consumption by using an opportunistic SU, which requires transmission power and frequency scanning time.
Another efficient energy management method from the first category proposed in [18] is based on the alternating optimization and dichotomy technique, which can optimise both the sensing time and the transmission power accordingly. The results show that the discussed approach outperforms the PSO algorithm, indicating that the practical integration of UAV and CR network is feasible.
In the second category, the researchers focused on spectrum hand-off as the main reason for power consumption. In [19,20], the authors identified spectrum hand-off, as the main factor for delays and energy losses. Thus, overuse of spectrum hand-offs is to be avoided. Furthermore, a spectrum hand-off approach is proposed based on the use of a delay violation ratio for CR networks. The given technique is implemented taking into account the unavailability of the current channel, channel characteristics and traffic characteristics, including a delay violation ratio based on the predefined packet delays. In addition, the elaborated calculation and simulation results illustrate that the proposed strategy can realize fewer hand-offs with an acceptable delay, i.e. it outperforms the other methods, which guarantees a lower probability of collision between SUs and significantly improves energy consumption. 
[21] investigated a collaborative compression-spectrum sensing technique applied in UAV schemes, using the UAV as a nodal point for distributed data collection. As a result, the decoding method mentioned above is considered adaptable in many scenarios, as it not only allows faster and more accurate detection of given targets, but also manages the interruption of wireless transmission and improves the energy efficiency system.
The authors in [22] have presented an approach for a circular UAV trajectory in which both flight radius and speed are optimised simultaneously to achieve maximum energy efficiency, taking into account the energy consumption of the propulsion system. The basic idea is to improve the energy consumption, but with a limitation in the trajectory planning. The numerical results obtained show that the proposed technique is significantly superior in terms of overall energy efficiency based on maximization and/or minimization.
On the other side, communication links are the main part of wireless communications and fifth-generation terrestrial (5G) is the natural way to provide network access. Since the UAV always flies in three-dimensional (3D) space and is located at a higher altitude, one of the most important links between the UAV and the terrestrial base station (BS) is the air-to-ground link (A2G) [23]. Due to the variability of A2G channels compared to ground communication (indoor and outdoor channels), channel transmission models are the crucial functions in the analysis of relay power allocation [23]. In [24] the authors proposed a highly efficient UAV-based cooperative, non-orthogonal A2G multiple access cooperative scheme (NOMA) to support cellular up-link transmissions. NOMA is a powerful approach for separating users in the power domain [25] to increase the correspondence rate of remote users (with lower receiving power) by allowing nearby users (with significant receiving power) to gain accessto information intended for remote users [26]. In [24], the authors compared the proposed solution with the traditional Orthogonal Frequency Division Multiple Access (OFDMA) scheme and the traditional ground-to-ground scheme NOMA (G2G). The results showed that the UAV achieved better performance with respect to spectral efficiency, energy efficiency, and cellular user access ratio using the proposed standard resource allocation scheme NOMA (RA). In another research study [27] the authors proposed an Energy-Efficient down-link between NOMA and Simultaneous Wireless Information and Power Transfer (SWIPT) with limitations throughput. The NOMA protocol was used to achieve the high spectral efficiency (SE) as it can provide a service to a multi user base at the same instance/policy/frequency but at different power levels, which results in a significant gain in spectral efficiency over trivial orthogonal multiple access [27,28]. Moreover, SWIPT technology is used to increase energy efficiency (EE). In practice, the process of information decoding and energy harvesting must be separated in SWIPT, which means that these two processes cannot operate in the same received signal to enable SWIPT. They used the genetic algorithm (GA) as an optimization solution for the power allocation ratio and the power allocation coefficients to make the EE below the average of the SE. The results of the metric SE -EE showed the maximum EE. The comparative view of communication based energy efficient management solutions is detailed in Table 1.
\begin{figure}
\centering
\includegraphics[scale=0.32]{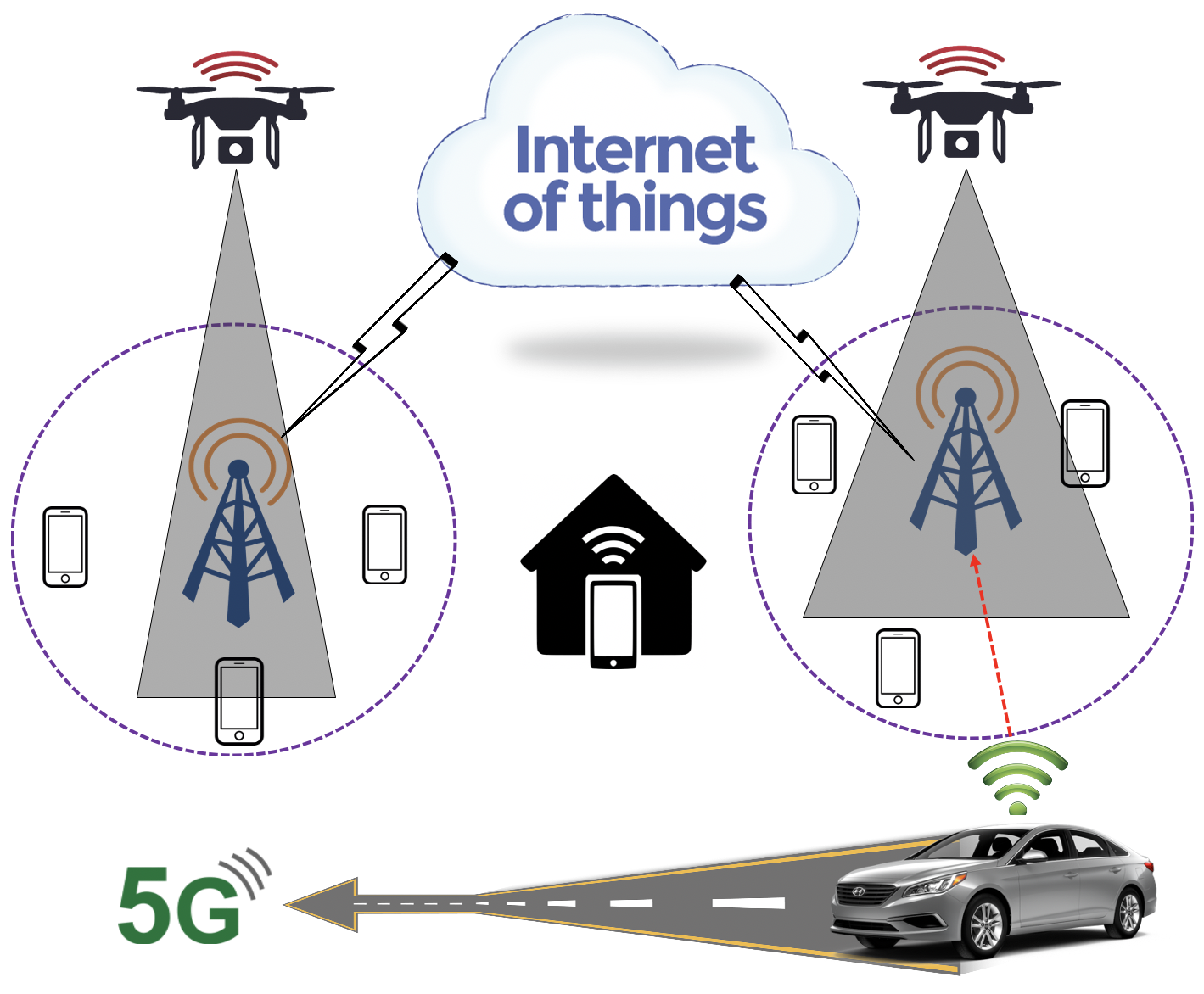}
\caption{Integration of UAVs within communications infrastructure through 5G networks}
\end{figure}
\begin{table*}[hb]
\caption{Energy-Efficiency (EE) Communication-based Techniques}
\label{tab=long table}
    \setlength\tabcolsep{3pt}
    \setcellgapes{3pt}
    \makegapedcells
\begin{tabular}{|
        >{$}p{\dimexpr0.15\linewidth-2\tabcolsep-1.33\arrayrulewidth}<{$}|
            p{\dimexpr0.33\linewidth-2\tabcolsep-1.33\arrayrulewidth}|
            p{\dimexpr0.16\linewidth-2\tabcolsep-1.33\arrayrulewidth}|
            p{\dimexpr0.31\linewidth-2\tabcolsep-1.33\arrayrulewidth}|
            p{\dimexpr0.06\linewidth-2\tabcolsep-1.33\arrayrulewidth}|}
    \hline
\thead{\text{Research area}}  &\thead{Issues / Objectives} &\thead{ Solutions} &\thead{Outcomes} &\thead{Ref.}
    \\  \hline
\text{\makecell{Energy optimization \\ of MUAVs}}  &   
\begin{itemize}
  \item Determine a joint optimized 3D location
  \item Optimal transmission power control 
  \item Exploiting the mobility of MUAVs
  \end{itemize}
  &  \centering{ A non-convex optimization \\
  (Weber formulation)}  & 
  \begin{itemize}
  \item Very close results to the PSO
  \item Better than other conceptional constructions
  \end{itemize}
 & [17] \\  \hline
\text{\makecell{UAV-based\\ overlay CR}} & 
\begin{itemize}
  \item Illustrated an efficient solution for energy management
  \item Improves the performance of the UAVs
  \item Optimization of secondary transmission power and sensing time 
  \end{itemize}
  & 
 \centering A Jointly optimization \\
 (Dichotomy technique)& 
  \begin{itemize}
  \item Significantly enhance the EE of the UAV-based CR
  \item Outperforms the PSO algorithm
  \end{itemize} &
  [18]    \\  \hline
\text{\makecell{Cognitive Radio\\ Networks (CRNs)}} & \begin{itemize}
  \item Reduce the unnecessary hand-offs 
  \item Considering a predefined packet delays 
  \end{itemize}
& \centering Delay violation ratio for CRNs & \begin{itemize}
  \item Less hand-offs 
  \item Less occupancy hand-off
  \item Acceptable packet delays
  \item Significantly improve the energy consumption
  \item Lower probability of collision between SUs
  \end{itemize} & [19,20]   \\  \hline
\text{\makecell{Transmission strategy\\ for CRs}} & 
The objective is to develop a strategy for CRs that can:
  \begin{itemize}
  \item Operate on a various PU's channels
  \item Help to proactive spectrum hand-off
  \end{itemize}  & \makecell{Optimal transmission \\with proactive\\ spectrum hand-off} & \begin{itemize}
  \item Minimum total cost 
  \item Highest data transmission efficiency
  \end{itemize} & [21,22]
\\  \hline
\text{Cooperative NOMA} & Improve the spectrum efficiency (SE) for the UAVs  & \centering Air-to-ground (A2G) cooperative with NOMA & \begin{itemize}
  \item Compared with G2G and OFDMA 
  \item Better performances in terms os SE and EE
   \item High access ratio of the CUs
  \end{itemize}& [24]    \\  \hline
\text{\makecell{EE down-link\\ for NOMA}} & Achieve high spectrum efficiency (SE) & \centering Down-link SWIPT-NOMA system & \begin{itemize}
  \item Compared based on SS-EE metrics 
  \item Achieve high EE at acceptable SE
  \end{itemize}& [27,28]    \\  \hline
\end{tabular}
\end{table*}
\begin{figure}
\centering
\includegraphics[scale=0.30]{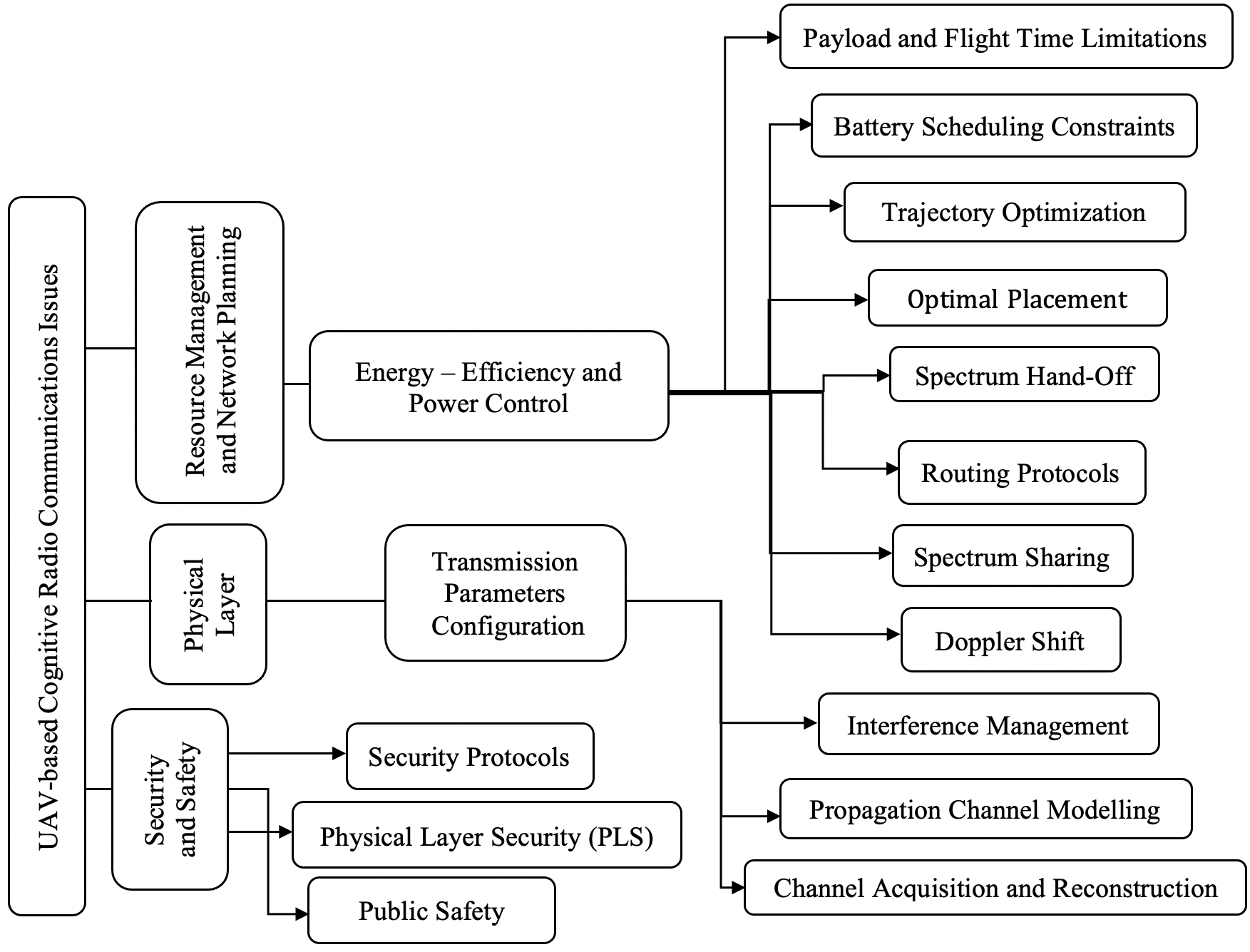}
\caption{Integration of UAV with CR: Important Challenges [13-26]} 
\end{figure}
\begin{table*}[hb]
\caption{Energy-Efficiency battery-based techniques}
\label{tab=long table}
    \setlength\tabcolsep{2pt}
    \setcellgapes{2pt}
    \makegapedcells
\begin{tabular}{|
        >{$}p{\dimexpr0.13\linewidth-2\tabcolsep-1.33\arrayrulewidth}<{$}|
            p{\dimexpr0.32\linewidth-2\tabcolsep-1.33\arrayrulewidth}|
            p{\dimexpr0.28\linewidth-2\tabcolsep-1.33\arrayrulewidth}|
            p{\dimexpr0.23\linewidth-2\tabcolsep-1.33\arrayrulewidth}|
            p{\dimexpr0.06\linewidth-2\tabcolsep-1.33\arrayrulewidth}|}
    \hline
\thead{\text{Research area}}  &\thead{Issues / Objectives} &\thead{Solutions} &\thead{Outcomes} &\thead{Ref.}
    \\  \hline

\text{Battery exchange} & Extending the flight endurance of UAV
  & A novel battery swapping technique & 
  The invention has proven to be effective & [31]    \\  \hline
 
 \text{\makecell{A mobile charging\\ infrastructure}}   &  Investigate several UAV recharging architectures &
 \begin{itemize} 
  \item UAV swapping
  \item Battery hot-swapping 
  \item Solar powered UAV
  \end{itemize} & 
  \begin{itemize} 
  \item Require backup UAV
  \item Noticeable downtime 
  \item Fails to ensure a continuous flight
  \end{itemize}  & 
  [29]    \\  \hline

\text{\makecell{Wireless power\\ transmission}} & Wireless charging in power-line based applications
 & Wireless energy transfer (WET) & A complete and reliable system does not exist yet & [32] \\  \hline

\text{\makecell{Solar powered\\ UAV}} & The core of this techniques is to design both:
\begin{itemize} 
  \item The corresponding objective functions in each stage 
  \item The switching conditions between adjacent stages
  \end{itemize}& 
  \begin{itemize} 
  \item Two types of three-dimensional (3D) flight based energy management strategies
  \item Both based on MPC framework
  \end{itemize}& 
  \begin{itemize} 
  \item The proposed strategy outperforms the traditional (2D) tracking
  \item Noticeable improvement under most wind conditions
  \end{itemize} &
  [33,34]    \\  \hline

\text{\makecell{Wireless far-field\\ charging}} &  Wireless charging the UAV battery  & This technique is utilized through:
\begin{itemize} 
  \item A multi-dimensional matrix of slot antennas 
  \item A self-supplementing bow-tie array implemented in each wing to receive rf energy.
  \item  Controlled by an MCU integration and MPPT algorithm implementation through a power stage the DC current from the arrays is processed
  \end{itemize}& 
  \begin{itemize} 
  \item Established a foundation for RF far-field charging
  \item Developed a functional prototyping method for the function model
  \end{itemize} & 
  [36]    \\  \hline

  \text{\makecell{NFV-based Energy\\ Scheduling}} & \begin{itemize}
  \item Construct the UAV scheduling scheme 
  \item Using brute-force search combination algorithm
  \end{itemize} & \centering NFV & Robustness and continuity of the service & [37]    \\  \hline

\text{Aerial base-station} & Leverage the DRL to enable energy efficient UAV control  & A highly energy efficient deep reinforcement learning DRL-based method & Significantly outperforms the two commonly used baseline methods (Greedy and Random) with respect to:
\begin{itemize} 
  \item Analysis
  \item Equity
  \item Energy consumption
  \end{itemize} & 
  [38]    \\  \hline

\end{tabular}
    \end{table*}

\section{UAV Battery Challenges and Potential Solutions}

In this section, we discussed the possible solutions, focusing on the battery limitations of UAVs. In general, battery-powered UAVs are susceptible to battery limitations because batteries are unreliable sources of energy that deviate from ideal behaviour during discharge [29]. Maintaining energy is critical for performing trivial tasks such as surveillance, spectrum sensing, search and rescue operations. Some techniques rely on mission planning, which depends on a pre-defined mission plan, by selecting an optimal path planning to ensure energy-efficient operation [30].
Overall, eliminating the problem of battery limitations would lead to a broader implementation of UAVs in many applications. To solve this problem, the inventor in [31] has invented a battery replacement technique that allows for extended flight time. This approach is performed by landing on an energy station, decoupling the first battery by a robotic arm, then connecting the first battery to a power source for recharging, and then connecting the second battery to the UAV.
Another solution proposed in [32] suggests charging the battery by wireless energy transmission (WET), especially in power line-based applications, by extracting the energy received from the wireless signal to maintain the UAV battery.
Recently, some research papers have presented renewable alternatives to take advantage of the environment [33]. [34], has investigated the use of a 3D energy management technology for a solar-powered UAV. In it, they explicitly address the unpredictable extended tracking missions that require longer flight times. The proposed strategy uses solar energy, potential gravitational force and wind to withstand different scenarios. The core of this method is the design of appropriate target functions and the ability to switch between different phases. The results are verified by comparing with 2D tracking, which proves that the presented technique is superior to conventional 2D tracking.
In [35] the authors designed an experimental solar-powered UAV that is capable of supplying the on-board electronics with power. The system is based on battery management by implementing Maximum Power Point Tracking (MPPT) and power conversion stages. MPPT is used to ensure that maximum power is drawn from the panel's solar cells. This technology offers a sub-optimal solution because solar energy is not ideal as it can only function during the day and is also susceptible to the unpredictable atmospheric changes that could lead to an inevitable failure. In [36], a wireless charging technology was proposed that would involve landing on a platform to recharge without changing batteries.
Moreover, a UAV scheduling scheme using the (NFV) Network Function Virtualization algorithm was discussed in [37]. The NFV is responsible for managing the entire process, including an energy management system that ensures continuity of service provided through the scheduling and replacement of the drones in service. Furthermore, this algorithm aims to ensure robust drone scheduling for a long period of time and in return maximise the use of available resources and provide a robust communication scheme that keeps the service running. Furthermore, energy-efficient drone control is investigated in [38] to achieve fair and effective communication coverage. The authors focused on addressing the limited range and resources as the main problems in UAV applications based on communication coverage. Therefore, they proposed a technique based on energy-efficient Deep Reinforcement Learning (DRL-EC3).The simulation results test and validate the system performance of the proposed (DRL-EC3) and show that (DRL-EC3) can achieve average energy efficiency compared to the basic random and greedy methods. The comparison between battery-based energy-efficient management solutions is presented in detail in Table 2.
\section{Future Road map}
One of the key design priorities for applications that integrate UAVs into communication-based systems is to extend up-time without compromising overall performance and sensing capabilities. However, today's massive growth in the communications industry, which requires the processing of large amounts of data and is complex in nature, can result in massive, irregular and time-varying data. In addition, down-sampling a time-varying signal in conventional signal processing, which results in compression of samples in the time domain, can negatively affect the performance of spectral sampling. Consequently, this leads to a noticeable loss of energy, which limits the effectiveness of UAVs. One of the potential tools for analysing these irregular data sets is Graph Signal Processing (GSP). GSP can extend conventional signal processing methods to accommodate time-varying signals on irregular structures such as graphs. In addition, GSP tools can aim to find out how to recover a graph signal from smaller nodes and what would be an optimal sampling subset if the possible bandwidth of the sampled graph signal were limited. This work can be planned by addressing goals such as finding the optimal power efficiency pattern based on the acquisition algorithm, low cost, and bandwidth traffic to achieve a robust energy efficient system.\\
Regarding the last section of battery-based efficiency, all batteries have a fixed capacity that decreases after some time, regardless of the technology and innovation used. Therefore, in narrowing down the state of the art, the problem of predicting the actual discharge time of the battery and the time in which the battery can no longer withstand flight conditions proves to be extremely crucial.
\section{Conclusions}
This paper examines both battery limitations and energy consumption issues through a detailed presentation and comparison of the latest techniques and algorithms developed by researchers and inventors, with a clear focus on the integration of UAVs in the communications sector, especially in CRN applications and battery-based energy management. Finally, we consider the implementation of the novel and emerging tool (Graph Signal Processing) as an opportunistic strategy to overcome the weaknesses of UAV communication networks and machine leaning techniques to address battery-based energy efficiency.


\end{document}